\documentclass[twocolumn,showpacs,prd,floatfix,axodraw]{revtex4}
\usepackage{mathrsfs}
\usepackage{graphicx,,booktabs,bm}
\usepackage{overpic}
\usepackage{color}
\usepackage{amssymb}
\usepackage{soul}

\usepackage{mathrsfs,bm,amsmath,amssymb}
\usepackage{longtable,lscape}
\usepackage{txfonts}
\usepackage{amssymb}
\usepackage{indentfirst}
\usepackage{graphicx,,booktabs}
\usepackage{multirow}
\usepackage{color}
\usepackage{amssymb}

\allowdisplaybreaks[4]

\definecolor{cover}{rgb}{0.77,0.87,0.88}
\definecolor{blueone}{rgb}{0.1,0.1,.7}
\definecolor{citec}{rgb}{0.14,0.47,0.09}
\definecolor{two}{rgb}{0.0,0.5,0.}
\definecolor{three}{rgb}{.5,.1,0.15}
\usepackage[bookmarks=true,bookmarksopen=false,plainpages=false,breaklinks=true,
   bookmarksnumbered=true,hypertexnames=false,
   filecolor=blue,urlcolor=three,menucolor=three,
   linkcolor=three,citecolor=blueone, colorlinks,
   anchorcolor=blue,runcolor=pink,frenchlinks=red
   pdfstartview=FitH,pdftitle=title,%
   pdfauthor=author]{hyperref}

\begin{document}
\title{Strong and radiative decays of $D\Xi$ molecular state and  newly observed $\Omega_c$ states}

\author{Yin Huang$^1$}
\thanks{huangy2017@buaa.edu.cn}
\author{Cheng-jian Xiao$^{3,6}$}
\author{Qi Fang L\"u$^{4}$}
\author{Rong Wang$^{5}$}
\author{Jun He$^{2}$}
\thanks{Corresponding author: junhe@njnu.edu.cn}
\author{Lisheng Geng$^1$}
\thanks{Corresponding author: lisheng.geng@buaa.edu.cn}

\affiliation{$^1$School of Physics and Nuclear Energy Engineering,
International Research Center for Nuclei and Particles in the Cosmos and
Beijing Key Laboratory of Advanced Nuclear Materials and Physics,
Beihang University, Beijing 100191, China\\
$^2$Department of Physics and Institute of Theoretical Physics, Nanjing Normal University,
Nanjing 210097, People¡¯s Republic of China\\
$^3$Institute of Modern Physics, Chinese Academy of Sciences, Lanzhou 730000, China\\
$^4$Synergetic Innovation Center for Quantum Effects and Applications
(SICQEA), Hunan Normal University, Changsha 410081, China\\
$^5$Institut de Physique Nucl\'eaire,
CNRS-IN2P3, Univ. Paris-Sud, Universit\'e Paris-Saclay,
91406 Orsay Cedex, France\\
$^6$University of Chinese Academy of Sciences, Beijing 100049, People¡¯s Republic of China}

\date{\today}
\begin{abstract}
In this work, we study  strong and radiative decays of  S-wave $D\Xi$ molecular state, which is related to the $\Omega^*_c$
states newly observed at LHCb.  The coupling between the $D\Xi$ molecular  state  and its constituents $D$ and $\Xi$ is calculated by using the
compositeness condition.  With the obtained coupling, the partial decay widths of the $D\Xi$ molecular state  into the $\Xi_c^{+}K^{-}$, $\Xi^{'+}_cK^{-}$
and $\Omega^{*}_c(2695)\gamma$ final states through hadronic loop are calculated with the help of  the effective Lagrangians.  By comparison with the LHCb observation,
the current results of total decay width  support  the $\Omega^{*}_c$(3119) or $\Omega^{*}_c$(3050) as  $D\Xi$ molecule  while the the decay width of the
$\Omega^{*}_c$(3000), $\Omega^{*}_c$(3066) and $\Omega^{*}_c$(3090) can not be well reproduced in the molecular state picture.  The partial decay widths are also
presented and helpful to further understand the internal structures of  $\Omega^{*}_c$(3119) and $\Omega^{*}_c$(3050).
\end{abstract}


\maketitle
\section{INTRODUCTION}
For a long time, little is known about the charmed baryon $\Omega_c$ with quantum numbers $C=1$ and $S=-2$, which is composed of one charm quark and two strange quark in the conventional constituent quark model. Only ground state $\Omega^{*}_c(2695)$ and $\Omega^{*}_c(2770)$ are listed in the newest version of
the Review of Particle Physics (PDG)~\cite{Patrignani:2016xqp}.   Recently, five new narrow $\Omega_c^{*}$
states named $\Omega^{*}_c(3000)$, $\Omega^{*}_c$(3050),$\Omega^{*}_c$(3066), $\Omega^{*}_c$(3090), and $\Omega^{*}_c$(3119)
were reported by the LHCb collaboration in the $\Xi_c^{+}K^{-}$ mass spectrum~\cite{Aaij:2017nav}.  Though the quantum numbers of
these new $\Omega_c^{*}$ states are not confirmed, it is very helpful to understand the charmed baryon spectrum.

The LHCb observation  stimulated a large amount of the theoretical  studies bout the new $\Omega_c^{*}$
states with different assumptions of their internal structures.  Naturally, many authors try to assign these states into the conventional  three-quark  frames. In Refs.~\cite{Karliner:2017kfm,Almasi:2017bhq,Chen:2016phw,Chen:2017gnu}
the new $\Omega^{*}_c$ baryons were interpreted as  $1P$ and $2S$ $\Omega^{*}_c$ baryons in the conventional quark models.
The QCD sum rules were also applied to study these states in three-quark picture~\cite{Wang:2017zjw,Agaev:2017jyt}. The lattice
calculation was also performed and try to determine their quantum numbers~\cite{Padmanath:2017lng}. In Refs.~\cite{Wang:2017hej,Chen:2017sci},
the authors investigated the decay properties to reveal the nature of these states.

It is quite rare to observe five states in one observation simultaneously. So many states observed also make it difficult to put all  states into the conventional quark model. Hence, after the observation at LHCb, the newly observed $\Omega_c^{*}$ was immediately interpreted as exotic  state beyond three-quark picture, $i.\ e.$, the pentaquark state.
The largest mass gaps between the newly observed $\Omega^{*}_c$ baryons and the ground $\Omega^{*}_c$ baryon are about 400 MeV,
which is large enough to excite  a light quark-antiquark pair.  Indeed, in Ref.~\cite{Yang:2017rpg}, pentaquark-like
$\Omega^{*}_c$ baryons were studied in the constituent quark model and   associated to some
of the LHCb $\Omega^{*}_c$ baryons.  In Ref.~\cite{An:2017lwg}, it was found that four $sscq\bar{q}$
states with $J^P=1/2^{-}$ or $J^P=3/2^{-}$  have  masses close to the newly observed $\Omega^{*}_c$
states.  In the chiral quark-soliton model,  pentaquark-like structures were suggested for
the $\Omega^{*}_c(3050)$ and $\Omega^{*}_c(3119)$~\cite{Kim:2017jpx,Kim:2017khv}. Since the $\Xi'_c \bar{K}$ and $\Xi D$
thresholds fall in the mass region of the LHCb observed $\Omega^{*}_c$ states,  hadronic molecule
interpretations can not be excluded. In Ref.~\cite{Montana:2017kjw},  the $\Omega^{*}_c$(3050) and $\Omega^{*}_c$(3090) were
regarded as meson-baryon molecules and with a similar method, the $\Omega^{*}_c$(3119) was also proposed to be a hadronic molecule~\cite{Debastiani:2017ewu}.
Moreover,  the $\Omega_c^{*}(3000)$, $\Omega^*(3050)$ ,and $\Omega^*(3090)$ or $\Omega^*(3119)$ can all be explained as meson-baryon molecular state in
Ref.~\cite{Nieves:2017jjx}.  With the one-gluon-exchange and the Goldstone-boson-exchange in addition to the color confinement, the authors
in Ref.~\cite{Huang:2017dwn} suggested that only  $\Omega^{*}_c$(3119) can be explained as an $S$-wave resonance state of $\Xi{}D$ with $J^P=1/2^{-}$,
which decays mainly through  $S$ wave into $\Xi_cK$ and $\Xi^{'}_cK$.

Until now, the nature of the observed $\Omega^{*}_c$ baryons remains unclear.  In addition to  their masses,  decay property
also serves as an important way to unveil  the nature of hadrons.   In Ref.~\cite{Wang:2017hej} the authors studied the decay
patterns of the $\Omega^{*}_c$ baryons  in a chiral quark model in  three-quark picture and  suggested that most of  the
low-lying $\Omega^{*}_c$ baryons have masses in the vicinity of the $\Xi_c^{+}K^{-}$ and $\Xi^{'+}_cK^{-}$ thresholds, to which
the strong decay will almost saturate their total decay widths.  However, the decays of the $\Omega^*_c$ baryons, which are helpful to understand their internal structures, have not been studied in the molecular state picture.

 In Refs.~\cite{Chen:2016qju,He:2015yva,Xiao:2016mho,Cleven:2013mka,He:2012zd},
the decays of hadronic molecular states have been studied by calculating the hadronic loop with the assumption that a molecular
state prefers to decay into its two constituents.  The  technique for evaluating composite hadron systems has been widely used
to study hadronic molecular state, where the compositeness condition, corresponding to $Z=0$, has been employed to extract the
coupling of a molecular state to its constituents~\cite{Dong:2008gb,Dong:2017rmg,Branz:2009yt,Faessler:2007gv}.   In this work, we will
calculate the radiative and strong decay pattern  of  S-wave $D\Xi$ molecular state
within the effective Lagrangians approach, and find the relation between the $D\Xi$ molecular state and the $\Omega_c^*$ states by comparing  with the LHCb observation,

This paper is organized as follows. The theoretical formalism is explained in Sec.~\ref{Sec: formulism}.
The predicted partial decay widths  are presented in Sec.~\ref{Sec: results}. Finally, we give  discussion and  summary in
the last section.

\section{FORMALISM AND INGREDIENTS}\label{Sec: formulism}
\begin{figure}[htbp]
\begin{center}
\includegraphics[scale=0.52]{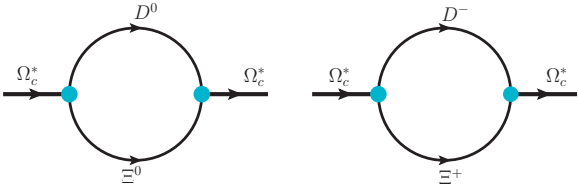}
\caption{Self-energy of the $\Omega_c^{*}$ states.}
\label{t-mass}
\end{center}
\end{figure}

In the molecule scenario, the interaction between the state $\Omega_c^{*}$ and its components $\Xi{}D$ is mainly via $S$-wave and the simplest
Feynman diagrams are shown in Figs.~\ref{t-mass}.   For the $\Omega_c^{*}\Xi{}D$ coupling, following Refs.~\cite{Dong:2008gb,Dong:2017rmg},  we take the Lagrangian densities as
\begin{align}
{\cal{L}}(x)=ig_{\Omega_c^{*}\Xi{}D}\Omega_c^{*}(x)&\int{}d^4y\Phi(y^2)[\bar{\Xi}^{0}(x+\omega_{D^0}y)D^{0}(x-\omega_{\Xi^0}y)\nonumber\\
            &+\bar{\Xi}^{+}(x+\omega_{D^-}y)D^{-}(x-\omega_{\Xi{^+}}y)]\label{eq1},
\end{align}
where $\omega_{D^{0,-}}=m_{D^{0,-}}/(m_{\Xi^{0,+}}+m_{D^{0,-}})$ and $\omega_{\Xi^{0,+}}=m_{\Xi^{0,+}}/(m_{\Xi^{0,+}}+m_{D^{0,-}})$.
In the Lagrangian, an effective correlation function $\Phi(y^2)$  is introduced to reflect the distribution of two constituents, $\Xi$ and $D$,  in the hadronic molecule $\Omega_c^{*}$ state. It also play a role to avoid the Feynman diagrams ultraviolet divergence, which requires that its Fourier
transform should vanish quickly in the ultraviolet region in the Euclidean space.  Since only S wave is considered  in  current work, we adopt an exponential  form  $\Phi(-p_E^2)\doteq{}\exp(-p_E^2/\chi^2)$
with $p_E$ being the Euclidean Jacobi momentum as used in Refs.~\cite{Dong:2008gb,Dong:2017rmg}.    The $\chi$ is a free size parameter characterizing the distribution of the two components in the molecule
and we adopt the $\chi=1$ that is often used in Refs.~\cite{Chen:2016qju,He:2015yva,Xiao:2016mho,Cleven:2013mka,Dong:2008gb,Dong:2017rmg,Branz:2009yt,Faessler:2007gv}  .

The only undetermined parameter is the coupling between molecular state and two constituents, $g_{\Omega_c^{*}\Xi{}D}$, which strength is a key factor to the value of the decay width on which we focus in the current work.  Following Refs.~\cite{Weinberg:1962hj,Salam:1962ap, Dong:2009tg}, we will adopt the compositeness condition to calculate the coupling of the hadronic molecule $\Omega_c^{*}$ and its consituents $\Xi$ and $D$.  This condition
requires that the renormalization constant of the hadronic molecular wave function is equal to zero,
$1-\frac{d\Sigma_{\Omega_c^{*}}}{dk\!\!\!/_0}|_{k\!\!\!/_0=m_{\Omega_c^{*}}}=0$,
with $\Sigma_{\Omega_c^{*}}$ being the self-energy of the hadronic molecule $\Omega_c^{*}$. Such relation connects the binding energy and the coupling strength of  bound state and its constituents. Now that the masses of $\Omega_c^*$ baryons have been observed in experiment, the couplings can be determined with such relation.

The Feynman diagram describing the self-energy of the $\Omega_c^{*}$ states is presented in
Fig.~\ref{t-mass}. With the help of the effective Lagrangian in Eq.~(\ref{eq1}), we can obtian the self energy of the $\Omega_c^*$ as
\begin{align}
\Sigma_{\Omega_c^{*}}(k_0)&=g^2_{\Omega_c^{*}\Xi{}D}\int\frac{d^4k_1}{(2\pi)^4i}\nonumber\\
                            &\times\{\Phi^2[(k_1-k_0\omega_{\Xi^{0}})^2]\frac{k\!\!\!/_1+m_{\Xi^0}}{k_1^2-m^2_{\Xi^0}}\frac{1}{(k_0-k_1)^2-m^2_{D^0}}\nonumber\\
                            &+\Phi^2[(k_1-k_0\omega_{\Xi^{+}})^2]\frac{k\!\!\!/_1+m_{\Xi^{+}}}{k_1^2-m^2_{\Xi^{+}}}\frac{1}{(k_0-k_1)^2-m^2_{D^{-}}}\},
\end{align}
where $k_0^2=m^2_{\Omega_c^{*}}$ with $k_0,m_{\Omega_c^{*}}$ denoting the four-momenta and
mass of the $\Omega_c^{*}$, respectively.  Here, we set $m_{\Omega_c^{*}}=m_{D}+m_{\Xi}-E_b$
with $E_b$ is the binding energy of $\Omega_c^{*}$.   While $k_1, m_{\Xi}$ and $m_D$ are the
four-momenta, mass of the $\Xi$ and mass of $D$, respectively.

According to the normalization conditions, the coupling constants is given by
\begin{align}
1/g^2_{\Omega_c^{*}\Xi{}D}=\int_0^{\infty}\frac{d\eta{}d\beta}{16\pi{}z^2}\sum_{i=1}^{2}H_ie^{-\frac{\omega_i}{\alpha^2}},~~~z=2+\eta+\beta\label{eq3}
\end{align}
with
\begin{align}
H_{1}&=\frac{2\omega_{\Xi^0}+\beta}{z}-\frac{2}{\alpha^2}[\frac{2\omega_{\Xi^{0}}+\beta}{z}m_{\Omega_c^{*}}+m_{\Xi^{0}}]\nonumber\\
                                   &\times{}[\frac{(2\omega_{\Xi^{0}}+\beta)^2}{z}+(2\omega^2_{\Xi^0}+\beta)]m_{\Omega_c^{*}}\\
\omega_1&=[\frac{(2\omega_{\Xi^0}+\beta)^2}{z}-2\omega^2_{\Xi^0}-\beta]m^2_{\Omega_c^{*}}+\eta{}m^2_{\Xi^0}+\beta{}m^2_{D^0},\\
H_{2}&=\frac{2\omega_{\Xi^+}+\beta}{z}-\frac{2}{\alpha^2}[\frac{2\omega_{\Xi^{+}}+\beta}{z}m_{\Omega_c^{*}}+m_{\Xi^{+}}]\nonumber\\
                                   &\times{}[\frac{(2\omega_{\Xi^{+}}+\beta)^2}{z}+(2\omega^2_{\Xi^+}+\beta)]m_{\Omega_c^{*}}\\
\omega_2&=[\frac{(2\omega_{\Xi^+}+\beta)^2}{z}-2\omega^2_{\Xi^+}-\beta]m^2_{\Omega_c^{*}}+\eta{}m^2_{\Xi^+}+\beta{}m^2_{D^-},
\end{align}
where the $\eta$ and $\beta$ will be integrated out, and the $\alpha$ is a free parameter, which will be discussed later.

Considering the quantum numbers and phase space, the strong decay modes of $\Omega^{*}_{c}$ are
$\Omega_c^{*}\to{}\Xi_c^{(')+}K^{-}$ and $\Omega_c^{*}\to{}\Xi_c^{(')0}\bar{K}^{0}$.
In this work, we only compute the partial decay width of  $\Omega_c^{*}\to{}\Xi_c^{(')+}
K^{-}$, and that of $\Omega_c^{*}\to{}\Xi_c^{(')0}\bar{K}^{0}$ can be obtained by  isospin symmetry
$\Gamma(\Omega_c^{*}\to{}\Xi_c^{(')+}K^{-})=\Gamma(\Omega_c^{*}\to{}\Xi_c^{(')0}\bar{K}^{0})$.
The sum of the two parts is the total decay width of the $\Omega_c^{*}\to{}\bar{K}\Xi_c^{(')+}$.

\begin{figure}[htbp]
\begin{center}
\includegraphics[scale=0.40]{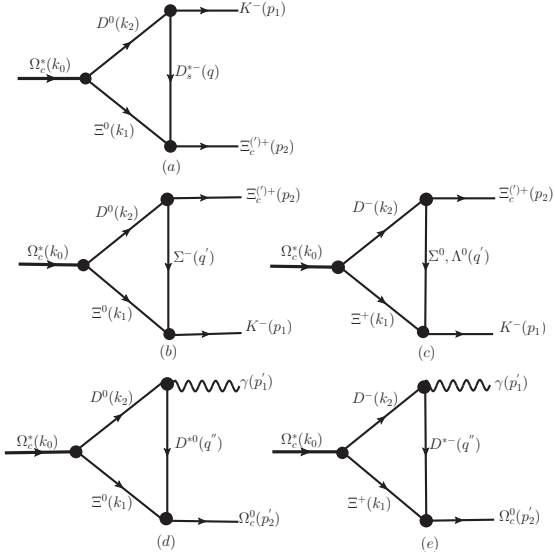}
\caption{Feynman diagrams for the $\Omega^{0*}_c\to{}K^{-}\Xi_c^{(')+}$ and $\Omega_c^{*}\gamma$ decay processes.}
\label{t-decay}
\end{center}
\end{figure}

In the hadronic molecule picture, $\Omega_c^{*}$ can decay into $\Xi_c^{+}K^{-}$, $\Xi^{'+}_cK^-$
and $\gamma\Omega^{*}_c(2695)$ by rearranging the quarks in its components.  At the hadron level,
$\Omega_c^{*}$ is treated as a bound state of $\Xi{}D$ and the decay $\Omega_c^{*}\to{}\Xi_c^{+}K^{-}$,
$\Xi^{'+}_cK^{-}$, and $\gamma\Omega^{*}_c(2695)$ occurs by exchanging a proper strange
meson and hyperon as shown in Fig.~\ref{t-decay}.  In the present work, we estimate these
triangle diagrams in an effective Lagrangian approach.  Besides the Lagrangian
in Eq.~\ref{eq1}, the effective Lagrangians of relevant interaction vertices
are also needed~\cite{Azevedo:2003qh,Hofmann:2005sw,Wang:2017smo,Xiao:2016hoa,Nakayama:2006ty}.
\begin{align}
&{\cal{L}}_{KDD_s^{*}}=ig_{KDD_s^{*}}D^{*\mu}_s(\bar{D}\partial_{\mu}\bar{K}-\bar{K}\partial_{\mu}\bar{D})+H.c.,\\
&{\cal{L}}_{\Xi^{0}D^{*-}_s\Xi^{+}_c}=\frac{\sqrt{6}g}{4}\bar{\Xi}^0\gamma_{\mu}D^{*-\mu}_s\Xi_c^{+}+H.c.,\\
&{\cal{L}}_{\Xi^{0}D^{*-}_s\Xi^{'+}_c}=\frac{\sqrt{2}g}{4}\bar{\Xi}^{0}\gamma_{\mu}D^{*-\mu}_s\Xi_c^{'+}+H.c.,\\
&{\cal{L}}_{\Xi{}D^{*}\Omega^0_c}=\frac{g}{\sqrt{2}}\bar{\Omega}^0_c\gamma^{\mu}D^{*+}_{\mu}\Xi^{-}+\frac{g}{\sqrt{2}}\bar{\Omega}^0_c\gamma^{\mu}D^{*0}_{\mu}\Xi^{0},\\
&{\cal{L}}_{\Xi_c^{(')}\Lambda{}D}=\frac{ig_{\Xi_c^{(')}\Lambda{}D}}{m_{\Xi_c^{(')}}+m_{D}}\bar{\Xi}_c\gamma^{\mu}\gamma^5\Lambda\partial^{\mu}\bar{D},\\
&{\cal{L}}_{\Xi_c^{(')}\Sigma{}D}=\frac{ig_{\Xi_c^{(')}\Sigma{}D^{}}}{m_{\Xi_c^{(')}}+m_{D}}\bar{\Xi}_c\gamma_{\mu}\gamma^5\vec{\tau}\cdot{}\vec{\Sigma}\partial^{\mu}\bar{D},\\
&{\cal{L}}_{\Xi\Lambda{}K}=\frac{ig_{\Xi{}\Lambda{}K}}{m_{\Lambda}+m_{\Xi}}\partial^{\mu}\bar{K}\bar{\Lambda}\gamma_{\mu}\gamma_5\Xi+H.c.,\\
&{\cal{L}}_{\Xi\Sigma{}K}=\frac{ig_{\Xi{}\Sigma{}K}}{m_{\Sigma}+m_{\Xi}}\partial^{\mu}\bar{K}\bar{\Sigma}\cdot\tau\gamma_{\mu}\gamma_5\Xi+H.c.,
\end{align}
where the $m_{\Lambda}$, $m_{\Sigma}$, and $m_{\Xi_c^{(')}}$ are the masses of the particle $\Lambda$, $\Sigma$,
and $\Xi_c^{(')}$, respectively.  The coupling constant $g_{KD_s^{*}D}=1.84$ is estimated in the framework of
light-cone QCD sum rules~\cite{Wang:2006ida} and $g=6.6$ is taken from~\cite{Hofmann:2005sw,Wang:2017smo}.
The $\vec{\tau}$ is the Pauli matrix, $\vec{\Sigma}$ represents the $\Sigma$ triplets,
and $\bar{D}$ and $\bar{K}$ are the doublets of charmed and $K$ mesons.
\begin{align}
\bar{D}=
\left(
  \begin{array}{c}
   D^{-} \\
    \bar{D}^{0} \\
  \end{array}
\right),~~~~~~~~~~
\bar{K}=
\left(
  \begin{array}{c}
  K^{-} \\
    \bar{K}^{0} \\
  \end{array}
\right).
\end{align}
The couplings for the different charge states are related by isospin symmetry:
\begin{align}
&\sqrt{2}g_{\Xi_c^{(')+}D^{0}\Sigma^{+}}=g_{\Xi_c^{(')+}D^{-}\Sigma^{0}}=\sqrt{2}g_{\Xi_c^{(')0}D^{-}\Sigma^{-}}=-g_{\Xi_c^{(')0}D^{0}\Sigma^{0}},\\
&\sqrt{2}g_{\Xi^{+}K^{0}\Sigma^{-}}=g_{\Xi^{+}K^{-}\Sigma^{0}}=\sqrt{2}g_{\Xi^{0}K^{-}\Sigma^{+}}=-g_{\Xi^{0}K^{0}\Sigma^{0}}.
\end{align}
One can estimate the couplings constants from SU(4) symmetry and phenomenological
constraints~\cite{Okubo:1975sc}
\begin{align}
&g_{\Xi_c\Sigma{}D}=\frac{3-2\alpha_{NN\pi}}{\sqrt{6}}g_{NN\pi},~~~~~~~~~g_{\Xi^{'}_c\Sigma{}D}=(2\alpha_{NN\pi}-1)g_{NN\pi},\nonumber\\
&g_{\Xi_c\Lambda{}D}=\frac{-3+2\alpha_{NN\pi}}{3\sqrt{2}}g_{NN\pi},~~~~g_{\Xi^{'}_c\Lambda{}D}=\sqrt{3}(2\alpha_{NN\pi}-1)g_{NN\pi},\nonumber\\
&g_{\Xi\Sigma{}K}=-g_{NN\pi},~~~~~~~~~~~~~~~~~g_{\Xi\Lambda{}K}=\frac{-4\alpha_{NN\pi}+3}{\sqrt{3}}g_{NN\pi},
\end{align}
where $g_{NN\pi}=13.26$~\cite{Nakayama:2006ty}, and $\alpha_{NN\pi}=0.64$~\cite{Okubo:1975sc}.
The numerical values of the couplings constant are listed in Table~\ref{table0}.
\begin{table}[h!]
\centering
\caption{Values of the effective meson-baryon couplings constants.}\label{table0}
\begin{tabular}{cccccc}
\hline\hline
~~~~~$g_{\Xi_c\Sigma{}D}$ & ~~~~~~$g_{\Xi^{'}_c\Sigma{}D}$ & ~~~~~~$g_{\Xi_c\Lambda{}D}$ &~~~~~~$g_{\Xi^{'}_c\Lambda{}D}$& ~~~~~~$g_{\Xi\Sigma{}K}$       & ~~~~~~$g_{\Xi\Lambda{}K}$                                   \\ \hline
~~~~~  9.31               & ~~~~~~ 3.71                    & ~~~~~~-5.38                 &~~~~~~6.43                      & ~~~~~~-13.26                  & ~~~~~~3.37                                                 \\  \hline
                     \hline
\end{tabular}
\end{table}

The involved interaction related to the photon field and the charmed mesons is ~\cite{Chen:2010re}.
\begin{align}
{\cal{L}}_{D^{*}D\gamma}&=\frac{g_{D^{*+}D^{+}\gamma}}{4}e\epsilon^{\mu\nu\alpha\beta}F_{\mu\nu}D^{*+}_{\alpha\beta}D^{-}\nonumber\\
                        &+\frac{g_{D^{*0}D^{0}\gamma}}{4}e\epsilon^{\mu\nu\alpha\beta}F_{\mu\nu}D^{*0}_{\alpha\beta}\bar{D}^{0}+H.c.,
\end{align}
where the field-strength tensors are defined as $F_{\mu\nu}=\partial_{\mu}A_{\nu}-\partial_{\nu}A_{\mu}$,
$D^{*}_{\alpha\beta}=\partial_{\alpha}D^{*}_{\beta}-\partial_{\beta}D^{*}_{\alpha}$, and $e=\sqrt{4\pi/137}$.
According to the Lagrangian and the radiative decay width of $\Gamma_{D^{*0}\to{}D^{0}\gamma}=26$ KeV that was deduced from the data on strong
and radiative decays of $D^{*}$ meson by theoretical predictions~\cite{Dong:2008gb,Chen:2015igx},  the coupling constant
$g_{D^{*0}D^0\gamma}$ can be determined as
\begin{align}
g_{D^{*0}D^0\gamma}=[\frac{96\pi{}m^3_{D^{*0}}}{e^2(m^2_{D^{*0}}-m^2_{D^0})^3}\Gamma_{D^{*0}\to{}D^{0}\gamma}]^{1/2}=2.0~{\rm GeV}^{-1}
\end{align}
where $m_{D^{*0}}=2.007$ GeV, $m_{D^0}=1.865$ GeV.   Similar, the coupling constant $g_{D^{*+}D^{+}\gamma}=-0.5$GeV$^{-1}$ is estimated from
the partial decay width of $\Gamma_{D^{*+}\to{}D^{+}\gamma}$=1.334 KeV~\cite{Patrignani:2016xqp} with $m_{D^{*\pm}}$=2.010 GeV.  The minus
sign is adopted according to the lattice QCD and QCD sum rule calculations~\cite{Becirevic:2009xp,Zhu:1996qy}.

In evaluating the amplitudes which are shown in Figs.~\ref{t-decay}, we need to include the form factors  because hadrons are not pointlike particles.
We adopt here the monopole-type form factor ${\cal{F}}_B(q^2)$ that was used in many previous works~\cite{Dong:2017rmg,Carvalho:2015eia},
\begin{align}
{\cal{F}}(q^2)=\frac{\Lambda^2-M^2}{\Lambda-q^2},
\end{align}
with $M$ being the mass of the exchanged meson and baryon. The cutoff $\Lambda=M+\lambda\Lambda_{QCD}$ with $\Lambda_{QCD}=220$ MeV is taken from Refs.~\cite{Huang:2013mua,Chen:2012nva}.
The parameter $\lambda$ reflects the nonperturbative property of QCD at the low-energy scale, which will be taken as a parameter and discussed later.

Putting all pieces together, we obtain the amplitudes for $\Omega_c^{*}(k_0)\to$ $[D\Xi]\to{}K^{-}\Xi_c^{(')+}$,
and $\Omega^{*}_c(2695)\gamma$ which correspond to the diagrams in Fig.~\ref{t-decay}, which reads
\begin{align}
{\cal{M}}_a&=-ig_{\Xi^{0}\Xi_c^{(')+}D^{*-}_s}g_{\Omega_c^{*}\Xi{}D}g_{K^{-}D^{0}D^{*-}_{s}}\int{}\frac{d^4q}{(2\pi)^4i}{\cal{F}}^2(q^2)\nonumber\\
           &\times\Phi((k_1\omega_{D^0}-k_2\omega_{\Xi^{0}})^2)\bar{u}(p_2)\gamma_{\mu}\frac{1}{k\!\!\!/_1-m_{\Xi^{0}}}u(k_0)\nonumber\\
           &\times{}(p_1^{\nu}+k_2^{\nu})\frac{-g_{\mu\nu}+q_{\mu}q_{\nu}/m^2_{D^{*-}_{s}}}{q^2-m^2_{D_{s}^{*-}}}\frac{1}{k_2^2-m^2_{D^{0}}},\\
{\cal{M}}_{b}&=\frac{-ig_{\Xi^{(')+}_c\Sigma^{-}D^0}g_{\Xi^0\Sigma^{-}K^-}g_{\Omega_c^{*0}\Xi{}D}}{(m_{\Xi_c^{(')+}}+m_{D^{0}})(m_{\Sigma^{-}}+m_{\Xi^{0}})}\int\frac{d^4q^{'}}{(2\pi)^4i}{\cal{F}}^2(q^{'2})\nonumber\\
             &\times\Phi((k_1\omega_{D^{0}}-k_2\omega_{\Xi^{0}})^2)\bar{u}(p_2)\gamma^{\mu}\gamma_{5}\frac{1}{q\!\!\!/^{'}-m_{\Sigma^-}}\gamma^{\nu}\gamma_5\nonumber\\
             &\times{}\frac{1}{k\!\!\!/_1-m_{\Xi^0}}u(k_0)\frac{1}{k_2^2-m^2_{D^0}}k_{2\mu}p_{1\nu},\\
{\cal{M}}^{\Lambda}_{c}&=\frac{-ig_{\Xi^{(')+}_c\Lambda{}D^{-}}g_{\Xi^{+}\Lambda{}K^{-}}g_{\Omega_c^{*0}\Xi{}D}}{(m_{\Xi_c^{(')+}}+m_{D^{-}})(m_{\Lambda}+m_{\Xi^{+}})}\int\frac{d^4q^{'}}{(2\pi)^4i}{\cal{F}}^2(q^{'2})\nonumber\\
             &\times\Phi((k_1\omega_{D^{-}}-k_2\omega_{\Xi^+})^2)\bar{u}(p_2)\gamma^{\mu}\gamma_{5}\frac{1}{q\!\!\!/^{'}-m_{\Lambda}}\gamma^{\mu}\gamma_5\nonumber\\
             &\times{}\frac{1}{k\!\!\!/_2-m_{\Xi^{+}}}u(k_0)\frac{1}{k_1^2-m^2_{D^{-}}}k_{2\mu}p_{1\nu},\\
{\cal{M}}_{d}&=\frac{egg_{\Omega_c^{0*}\Xi{}D}g_{D^{*0}D^{0}\gamma}}{4\sqrt{2}}\int\frac{d^4q^{''}}{(2\pi)^4i}{\cal{F}}^2(q^{''2})\nonumber\\
             &\times\Phi((k_1\omega_{D^0}-k_2\omega_{\Xi^{0}})^2)\bar{u}(p_2)\gamma_{\rho}\frac{1}{k\!\!\!/_1-m_{\Xi^0}}u(k_0)\nonumber\\
             &\times{}\epsilon^{\mu\nu\alpha\beta}(g_{\nu\eta}p^{'}_{1\mu}-g_{\mu\eta}p^{'}_{1\nu})(g_{\beta\sigma}q^{''}_{\alpha}-g_{\alpha\sigma}q^{''}_{\beta})\varepsilon^{*\eta}\nonumber\\
             &\times\frac{-g_{\rho\sigma}+q^{''}_{\rho}q^{''}_{\sigma}/m^2_{D^{*0}}}{q^{''2}-m^2_{D^{*0}}}\frac{1}{k_2^2-m^2_{D^0}}.\\
{\cal{M}}_{e}&=\frac{egg_{\Omega_c^{0*}\Xi{}D}g_{D^{*+}D^{-}\gamma}}{4\sqrt{2}}\int\frac{d^4q^{''}}{(2\pi)^4i}{\cal{F}}^2(q^{''2})\nonumber\\
             &\times\Phi((k_1\omega_{D^{-}}-k_2\omega_{\Xi^{+}})^2)\bar{u}(p_2)\gamma_{\rho}\frac{1}{k\!\!\!/_1-m_{\Xi^+}}u(k_0)\nonumber\\
             &\times{}\epsilon^{\mu\nu\alpha\beta}(g_{\nu\eta}p^{'}_{1\mu}-g_{\mu\eta}p^{'}_{1\nu})(g_{\beta\sigma}q^{''}_{\alpha}-g_{\alpha\sigma}q^{''}_{\beta})\varepsilon^{*\eta}\nonumber\\
             &\times\frac{-g_{\rho\sigma}+q^{''}_{\rho}q^{''}_{\sigma}/m^2_{D^{*-}}}{q^{''2}-m^2_{D^{*-}}}\frac{1}{k_2^2-m^2_{D^-}}.
\end{align}

The corresponding partial decay widths then read
\begin{align}
\Gamma[\Omega_c^{*}\to{}K^{-}\Xi_c^{(')+},\gamma\Omega^{*}_c(2695)]=\frac{1}{2J+1}\frac{1}{8\pi}\frac{|\vec{p}_{1}^{K/\gamma}|}{m^2_{\Omega_c^{*}}}\overline{|{\cal{M}}|^2},
\end{align}
where $J$ is the total angular momentum of the initial $\Omega_c^{*}$ state, the overline indicates the sum over the polarization vectors of final hadrons.  Here
$|\vec{p}^{K/\gamma}_1|$
is the 3-momenta of the decay products in the center of mass frame.

\section{RESULTS}\label{Sec: results}
\begin{figure}[h!]
\begin{center}
\includegraphics[scale=0.3]{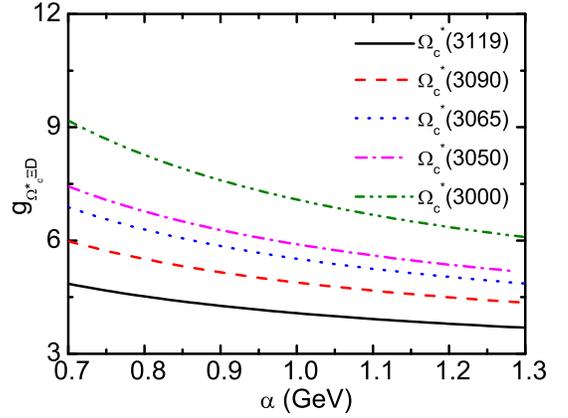}
\caption{(Color online) Coupling constants, $g_{\Omega_c^{*}\Xi{}D}$ (GeV$^{-1}$), for different $\Omega_c^{*}$
states as a function of the parameter $\alpha$.}
\label{cc}
\end{center}
\end{figure}

Regarding the five new $\Omega_c^{*}$ as $\Xi{}D$ hadronic molecules,  the coupling constants $g_{\Omega_c^{*}\Xi{}D}$
can be estimated from the compositeness condition. As shown in Eq.~(\ref{eq3}), the coupling constant is dependent on the parameter $\alpha$. In Fig.~\ref{cc}, we show the dependence of the coupling constants $g_{\Omega_c^{*}\Xi{}D}$
on the cutoff parameter $\alpha$.  The coupling constant $g_{\Omega_c^{*}\Xi{}D}$ decreases with the increase of $\alpha$. Taking the $\Omega^{*}_c(3119)$ as an example,    the value of the coupling constant  $g_{\Omega_c^*(3119)\Xi{}D}$ is not very sensitive to the model parameter $\alpha$ when varying cutoff parameter $\lambda$ from 0.7 GeV to 1.3 GeV (not sensitive to $\lambda$ also).   Fixing the $\alpha$ at certain value, such as $1.00$ GeV,  the coupling constants decrease with increase of  the mass $m_{\Omega_c^{*}}$.
According to the studies of the XYZ resonances and  the deuteron~\cite{Chen:2015igx,Dong:2014zka},
a typical value of $\alpha\sim$1 GeV is often employed.  Thus, in this work we take $\alpha=1.0$ and the corresponding
coupling constants are listed in Table.~\ref{table1}, which are used to calculate the decay processes of Fig.~\ref{t-decay}.
\begin{table}[h!]
\centering
\caption{Coupling constants, $g_{\Omega_c^{*}\Xi{}D}$, for different $\Omega_c^*$ states with
$\alpha=1$ GeV.}\label{table1}
\begin{tabular}{cccccc}
\hline\hline
   &$\Omega_c^{*}(3000)$~~  &$\Omega_c^{*}(3050)$~~&$\Omega_c^{*}(3065)$~~&$\Omega_c^{*}(3090)$~~&$\Omega_c^{*}(3119)$\\ \hline
$g_{\Omega^{*}_c\Xi{}D}$ &7.08              ~~  &5.90~~              &5.52             ~~&4.89             ~~&4.08\\               \hline\hline
\end{tabular}
\end{table}

\begin{figure}[h!]
\begin{center}
\includegraphics[bb=50 0 1000 635, clip,scale=0.4]{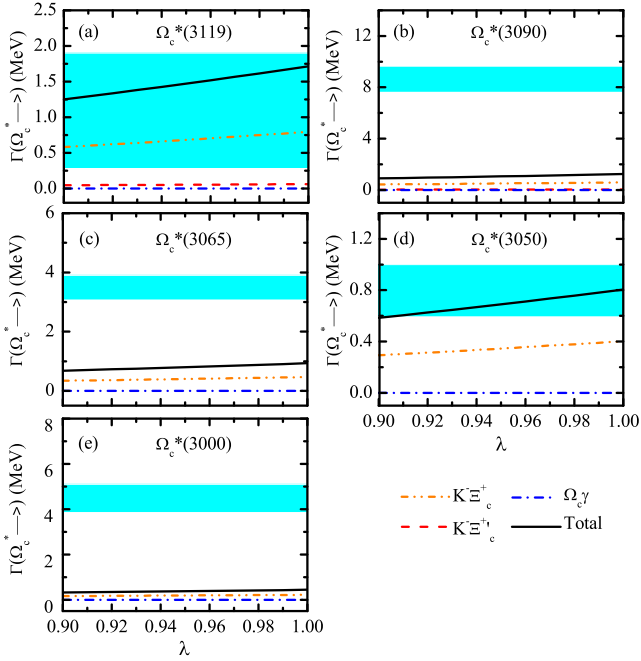}
\caption{(Color online) Partial decay widths of the $\Omega_c^{*}\to{}K^{-}\Xi^{+}_c$ (Orange dash dot dot line), $\Omega_c^{*}\to{}K^{-}\Xi_c^{'+}$ (red dash line), $\Omega_c^{*}\to{}\gamma\Omega_c(2695)$(blue dash dot),
and the total decay width (black solid line) with different $\Omega^{*}_c$
states depending on the parameter $\lambda$.  The oycn error bands correspond to the
total width observed in experiment.\cite{Aaij:2017nav}.}
\label{decaywidth}
\end{center}
\end{figure}

Once the coupling constants of the molecular $\Omega^*_c$ baryons and  $\Xi D$ are determined,
the partial decay widths of the $\Omega_c^{*}\to{}\Xi^{+}_cK^{-}$, $\Omega_c^{*}\to{}\Xi^{'+}_cK^{-}$,
$\Omega_c^{*}\to{}\Omega^{*}_c(2695)\gamma$, and the total decay width are  only dependent on the parameter $\lambda$ in the cutoff.
Though the value of $\lambda$  could not be determined in first principles, it is usually chosen as about 1 in the literature.  In Ref.~\cite{Dong:2017rmg}, by comparing the sum of the partial decay modes of the $\eta(2225)$
and $\phi(2170)$ with the total width, the parameter $\lambda$ was constrained as $\lambda=0.91-1.00$.   In addition, the experimental branching ratios of
$\psi(4040)\to{J/\psi\eta}$ and $\psi(4160)\to{J/\psi\eta}$ can be well explained with $\lambda=0.53-1.20$~\cite{Chen:2012nva}. Larger range of  0.5 to 5 can be found in Refs.~\cite{Liu:2006df,Colangelo:2002mj,Meng:2007cx,Liu:2008tv}.  Considering the values adopted in above literatures,
we adopt a parameter $\lambda$  in the a range of  $0.91 \leq \lambda \leq 1.0$ because
this range is determined from the experimental data of branching ratios within the same theoretical framework adopted in current work in Ref.~\cite{Dong:2017rmg}.
The numerical results are presented in
Fig.~\ref{decaywidth} with the variation of $\lambda$ from 0.90 to 1.0. In the discussed range,
the partial decay widths increase with $\lambda$, and the $\Omega_c^*$ states mainly decay
into $\Xi_c\bar{K}$ and the partial width into $\Xi_c{}\bar{K}$ is much larger than those into
$\Xi_c^{'}\bar{K}$ and, of course, $\gamma{}\Omega^{*}(2695)$.  The total width of $\Omega_c^*(3119)$ and $\Omega^*_c(3050)$ can be well reproduced in the $\lambda$ range considered here.  If we increase $\lambda$ to higher values, the total widths of all five $\Omega^*_c$ baryons can not be reproduced until a much larger $\lambda$ value of about 2 adopted. Hence,  it is reasonable to adopt a $\lambda$ of about 1 in the current work.

\begin{figure}[h!]
\begin{center}
\includegraphics[bb=25 30 600 410, clip,scale=0.46]{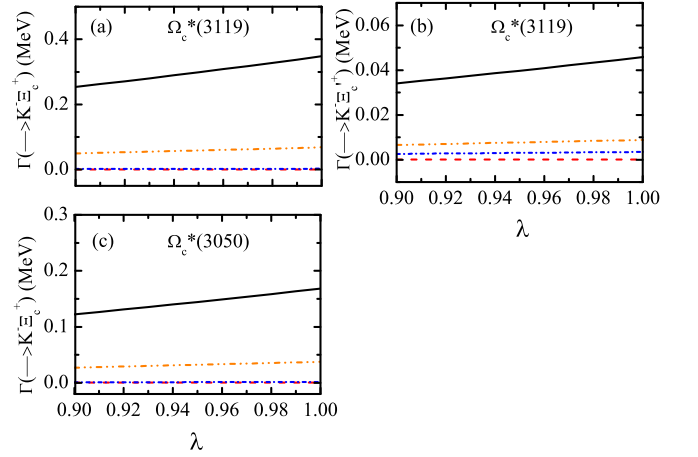}
\caption{(Color online) Individual contributions of the $D_s^{*-}$, $\Lambda$, $\Sigma^{-}$, and $\Sigma^{0}$
exchange for the $\Omega_c^{*}\to{}K^{-}\Xi_c^+$ and $\Omega_c^{*}\to{}K^{-}\Xi_c^{'+}$ for $\Omega^{*}_c$(3119) and
$\Omega^{*}_c$(3050) depending on the parameter $\lambda$.  The red dash, blue dash dot, black solid, and orange dash dot dot lines
stand for the $D_s^{*-}$, $\Lambda$, $\Sigma^{-}$, and $\Sigma^{0}$ contributions, respectively.}
\label{indivialwidth}
\end{center}
\end{figure}

The individual contributions of the $D_s^{*-}$, $\Lambda$, $\Sigma^{-}$, and $\Sigma^{0}$
exchanges for the decays $\Omega_c^{*}\to{}K^{-}\Xi_c^+$ and $\Omega_c^{*}\to{}K^{-}\Xi_c^{'+}$ are calculated and
presented in Fig.~\ref{indivialwidth}.   Since the relative signs of the three Feynman diagrams
[(a),(b),(c) in Fig.~\ref{t-decay}] are well defined,   the total decay widths obtained are the
square of their coherent sum.  It is found that the $\Sigma^{-}$ exchange plays
a dominant role,  while the $D_s^{*-}$, $\Lambda$, and $\Sigma^{0}$ exchanges give minor contributions.
However, the interferences among them are still sizable.  Even for the $\Omega_c^{*}$(3119) case,
the $|{\cal{M}}_{D^{*-}}+{\cal{M}}_{\Lambda}+{\cal{M}}_{\Sigma^{0}}|$ is about 1/8$\sim1/3$ of the
${\cal{M}}_{\Sigma^{-}}$, which contributes to the decay constructively.

The five new $\Omega_c^{*}$ particles were observed as resonances in the $\Xi_c^{+}K^{-}$
invariant mass distribution and are in the vicinity of the $\Xi_c\bar{K}$ and $\Xi^{'}_c\bar{K}$ thresholds.  The transition
$\Omega_c^{*}\to{}\Xi_c\bar{K}$ and $\Omega_c^{*}\to{}\Xi^{'}_c\bar{K}$ may be considered as  main
decay channels, the sum of which almost saturates the total width of each $\Omega_c^{*}$.  For the $\Omega^{*}_c(3090)$,
$\Omega_c^{*}(3000)$, and $\Omega_c^{*}(3065)$, their total decay widths are much smaller than the experimental total width.
Such results disfavor the assignment of these three states as $D\Xi$ molecular state.    Hence,
only the $\Omega_c^{*}$(3119) or $\Omega_c^{*}$(3050) states can be considered as $S-$wave $\Xi{}D$
molecules.  Hence, we only list the decay widths of $\Omega_c^{*}$(3119) and
$\Omega_c^{*}$(3050) with $\lambda=0.91-1.00$  in
Table.~\ref{tablewidth}.  For comparisons,  we show the results in the constituent quark model as
well~\cite{Wang:2017hej}. The decay width $\Gamma_{\Xi_c\bar{K}}$ is close to that in the constituent quark model  if we assign  the $S$-wave $\Xi D$ bound state as $\Omega_c^*(3050)$. Assuming this channel is dominant decay channel, the total decay width under such assignment is also consistent with that in constituent quark model and the experimental value.  Under assignment  as  $\Omega^*_c(3119)$,  the total widths decay width $\Gamma_{\Xi_c\bar{K}}$  is lager than that in the constituent  quark model while
$\Gamma_{\Xi'_c\bar{K}}$ is smaller, which leads to a comparable total decay width to  those in the constituent quark model and in experiment.

\begin{table*}[htbp!]
\centering
\caption{ Partial decay widths of $\Omega_c^{*}\to{}\Xi_c\bar{K}$,$\Xi_c^{'}\bar{K}$,$\Omega^{*}_c(2695)\gamma$, and the total decay width $\Gamma_{total}$ with
$\lambda=0.91-1.00$ that is introduced by the form factor, in comparison with the results in  the constituent quark model~\cite{Wang:2017hej}.  The total width
obtained from the LHCb experiments~\cite{Aaij:2017nav}.  All masses and widths are in units of MeV.  The two values of decay width for the $\Omega^*_c(3119)$ in Ref~\cite{Wang:2017hej} are for the assignments $|2^2S_{\Lambda\Lambda}1/2^{+}\rangle$ or $|2^4S_{\Lambda\Lambda}3/2^{+}\rangle$, respectively.
}\label{tablewidth}
\begin{tabular}{c|c|c|c|c|c|c|c|c|c}
\hline\hline
state&\multicolumn{2}{c|}{$\Gamma_{\Xi_c\bar{K}}$} &\multicolumn{2}{c|}{$\Gamma_{\Xi_c^{'}\bar{K}}$}&\multicolumn{2}{c|}{$\Gamma_{\Omega^{*}_c(2695)\gamma}$} &\multicolumn{3}{c}{$\Gamma_{total}$} \\ \hline
& This work     &Ref.\cite{Wang:2017hej}&~This work ~   &~Ref.\cite{Wang:2017hej}~ &This work($\times10^{-7}$) &Ref.\cite{Wang:2017hej}($\times10^{-3}$) &This work&Ref.\cite{Wang:2017hej} & Exp.\cite{Aaij:2017nav}           \\ \hline
$\Omega^{*}_c(3050)$ & $0.61-0.81 $ &0.61     & $...$         &$...$                             &$1.06-1.42$&$1.12   $              &0.61-0.81 &0.94                            &~~~$0.8\pm0.2 $          \\
$\Omega^{*}_c(3119)$ & $1.20-1.59 $ &0.6/0.6  & $0.094-0.122$ &0.45/0.11                         &$1.85-2.48$&$2.9/1.0$              &1.29-1.71 &1.15/0.73                       &~~~$1.1\pm0.8 $          \\
\hline\hline
\end{tabular}
\end{table*}

\begin{figure}[h!]
\begin{center}
\includegraphics[bb=22 15 600 260, clip,scale=0.47]{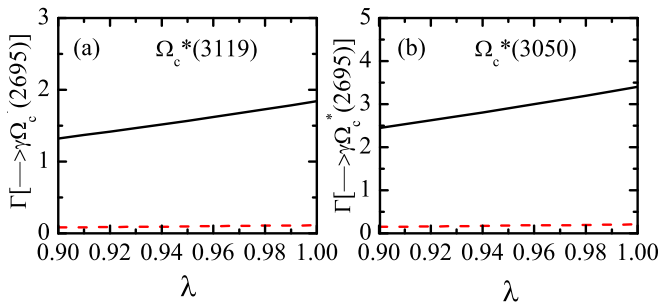}
\caption{(Color online) Individual contributions of the $D^{*0}$ and $D^{*-}$ exchange for the
$\Omega_c^{*}\to{}\Omega_c(2695)\gamma$ for different $\Omega^{*}_c$ states as a function of the parameter $\lambda$.
The black solid and red dash lines stand for the $D^{*0}$ and $D^{*-}$ contributions, respectively.
The numerical results are in units of $10^{-7}$ MeV.}
\label{indivial-photon}
\end{center}
\end{figure}

Now we turn to  the radiative decay $\Omega_c^{*}\to\Omega^{*}_c(2695)\gamma$.
The individual contributions of the $D^{*0}$ and $D^{*-}$ exchange and total decay width with varying $\lambda$
from 0.90-1.00 for the $\Omega_c^{*}\to{}\Omega^{*}_c(2695)\gamma$ are presented in Fig.~\ref{indivial-photon} and
Fig.~\ref{decaywidth}, respectively.  Our study shows that the partial width of the
$\Omega_c^{*}\to{}\Omega^{*}_c(2695)\gamma$ is rather small and the  $D^{*0}$ exchange
plays a dominant role, weakly increasing with the $\lambda$ increasing.  In the considered parameter region, the partial
widths for the $\Omega_c^{*}\to{}\Omega^{*}_c(2695)\gamma$ are predicted and listed in
Table.~\ref{tablewidth}, compared with the results in conventional charmed baryons scheme~\cite{Wang:2017hej}.
The partial widths of $\Omega_c^{*}(3119)\to{}\Omega^{*}_c(2695)\gamma$ and $\Omega_c^{*}(3050)\to{}\Omega^{*}_c(2695)\gamma$
in Ref.~\cite{Wang:2017hej} were 1.2 and 2.9/1.0$\times10^{-3}$ MeV, respectively, which are totally different
with the results in the present work.

\section{Discussion and SUMMARY}

In this work, the S-wave $D\Xi_c$ molecular states were studied by calculating their strong and radiative decays
to investigate whether the five new narrow $\Omega_c^{*}$ baryons, $\Omega^{}_c(3000)$, $\Omega_c^{*}$(3050),$\Omega_c^{*}$(3066),
$\Omega_c^{*}$(3090), and $\Omega_c^{*}$(3119), can be understood as  $\Xi{}D$ molecules.  With the coupling constants
obtained by composition condition, the decays through hadronic loops are calculated in a phenomenological effective Lagrangian
approach.The total decay widths can be well reproduced with the assumption  that the $\Omega_c^{*}$(3119) or $\Omega_c^{*}$(3050) as
$S$-wave $\Xi{}D$ bound state with $J^p=1/2^{-}$, which decay channels are $S-$wave $\Xi_cK$,
$\Xi^{'}_cK$ and $\Omega^{*}_c(2659)\gamma$.  The other newly reported $\Omega_c^{*}$ states cannot be
accommodated in the current molecular picture.  If the $\Omega_c^{*}$(3119) or $\Omega_c^{*}$(3050) is pure $D\Xi$
molecule, the radiative transition strength is quite small and the decay width is of the order of about 0.1 eV.

It is  interesting to compare our results with those in Refs.\cite{Wang:2017hej,Kim:2017jpx,Kim:2017khv,Nieves:2017jjx,Montana:2017kjw,Debastiani:2017ewu,Huang:2017dwn}.
According to Ref.~\cite{Wang:2017hej}, the $\Omega^*_c$ baryons at LHCb may be
conventional charmed baryons with $P-$wave or even higher partial waves.
In Refs.~\cite{Montana:2017kjw,Debastiani:2017ewu} the $J^p=1/2^{-}$  state is identified as a
meson-baryon molecule that can be associated to  the $\Omega_c^{*}$(3050), in agreement with our
conclusion.  However they claim that the $\Omega_c^{*}$(3050) only has a small $D\Xi$ component.
In Ref.~\cite{Debastiani:2017ewu} a $J^p=3/2^{-}$ state  can be associated
to the experimental $\Omega_c^{*}$(3119).  This is quite different from the our results
and those by Huang et al.~\cite{Huang:2017dwn}  that the $\Omega_c^{*}$(3119) can be explained as $S$-wave $\Xi{}D$ state
  with $J^p=1/2^{-}$.  Furthermore, in the chiral quark-soliton model,  pentaquark-like structures
were suggested for the $\Omega_c(3050)$ and $\Omega_c(3119)$~\cite{Kim:2017jpx,Kim:2017khv}.
It is very interesting to find that authors in the Ref.~\cite{Nieves:2017jjx} regarded $\Omega_c(3050)$
and $\Omega^{*}_c(3090)$ or $\Omega^{*}_c(3119)$ as meson-baryons.   However, a completely different conclusion was
drawn from Refs.~\cite{Wang:2017hej}  that the $\Omega_c^{*}$(3119) and $\Omega_c^{*}$(3050) can be understood
as conventional $css$ states. The radiative decay $\Omega_c^{*}\to\Omega^{*}_c(2695)\gamma$  may be helpful to distinguish these results.
If the $\Omega_c^{*}$(3119) or $\Omega_c^{*}$(3050) is pure $D\Xi$ molecule, the radiative transition
strength is quite small and the decay width is of the order of about 0.1 eV.  Future experimental
measurements of such a process can be quite useful to test the different interpretations of the
$\Omega_c^*(3119)$ and $\Omega_c^*(3050)$.

\section*{Acknowledgments}
This work is partly supported by the National Natural Science Foundation of China under Grants No.11522539, 11675228,11735003
and the fundamental Research Funds for the Central Universities.

\end{document}